\documentclass[prl,aps,twocolumn,showpacs,floatfix]{revtex4}%
\usepackage{amsmath}
\usepackage{amsfonts}
\usepackage{amssymb}
\usepackage{graphicx} 
\usepackage{bm}
 
\begin{document}


\title{Unconventional London penetration depth in Ba(Fe$_{0.93}$Co$_{0.07}$)$_2$As$_2$ single crystals}

\author{R.~T.~Gordon}
\author{N.~Ni}
\author{C.~Martin}
\author{M.~A.~Tanatar}
\author{M.~D.~Vannette}
\author{H.~Kim}
\author{G.~Samolyuk}
\author{J.~Schmalian}
\author{S.~Nandi}
\author{A.~Kreyssig}
\author{A.~I.~Goldman}
\author{J.~Q.~Yan}
\author{S.~L.~Bud'ko}
\author{P.~C.~Canfield}
\author{R.~Prozorov}
\email[corresponding author: ]{prozorov@ameslab.gov}
\affiliation{Ames Laboratory and Department of Physics \& Astronomy, Iowa State University, Ames, IA 50011}

\date{11 October 2008}
 
 \begin{abstract}
The London penetration depth, $\lambda(T)$, has been measured in several single crystals of Ba(Fe$_{0.93}$Co$_{0.07}$)$_2$As$_2$. Thermodynamic, electromagnetic, and structural characterization measurements confirm that these crystals are of excellent quality. The observed low temperature variation of $\lambda(T)$ follows a power-law, $\Delta \lambda (T) \sim T^n$ with $n=2.4 \pm 0.1$, indicating the existence of normal quasiparticles down to at least $0.02T_c$. This is in contrast to recent penetration depth measurements on single crystals of NdFeAsO$_{1-x}$F$_x$  and SmFeAsO$_{1-x}$F$_x$, which indicate an anisotropic but nodeless gap. We propose that a more three-dimensional character in the electronic structure of Ba(Fe$_{0.93}$Co$_{0.07}$)$_2$As$_2$  may lead to an anisotropic $s-$wave gap with point nodes that would explain the observed $\lambda(T)$. 
\end{abstract}


\maketitle
The discovery of superconductivity in LaFeAs(O$_{1-x}$F$_x$), with T$_c \approx 23$ K \cite{Kamihara08}, and T$_c$ above 50 K in SmFeAs(O$_{1-x}$F$_{x}$) \cite{Zhi-An08} has lead to a flurry of experimental and theoretical activities aimed to understand the fundamental physics governing this new family of superconductors. Much effort has been put forth to characterize the oxypnictide compounds RFeAsO$_{1-x}$F${_y}$ (R=rare earth, ``1111'' in the following), as well as oxygen-free A(Fe$_{1-x}$T$_x$)$_2$As$_2$ (A = alkaline earth and T = transition metal, ``122'' in the following). The major questions yet to be answered are the symmetry of the superconducting gap and its universality among all Fe-based pnictides. High transition temperatures, high upper critical fields, and the existence of a unique structural element, Fe-As layers, prompt a comparison with the high-T$_c$ cuprates.

It has now been well established that the cuprates exhibit d-wave symmetry of the order parameter \cite{cuprates_review}. In Fe-based pnictides the debate is wide open. Using different techniques, several groups have arrived at different conclusions, such as gapped versus nodal, or single versus multigap superconducting states. Tunneling measurements showed an unconventional order parameter with nodes \cite{Samuely2008}, two superconducting gaps and a pseuodogap \cite{Szabo2008}, as well as conventional s-wave BCS behavior \cite{Chen2008}. Measurements of specific heat have also shown both conventional \cite{Mu2008} and unconventional \cite{Mu2008a} behavior. Angle-resolved photoemission (ARPES) data are still in disagreement regarding the gap anisotropy, amplitude, and comparison to theoretical electronic structure calculations \cite{Liu08,Evtushinsky2008}.

As far as penetration depth is concerned, measurements reported for the 1111 system are consistent with regard to the overall gap topology. Microwave cavity perturbation in PrFeAsO$_{1-x}$ \cite{Hashimoto2008} and muon spin relaxation ($\mu$SR) in both LaFeAsO$_{1-x}$ \cite{Luetkens2008} and SmFeAsO$_{1-x}$F$_x$  \cite{Drew2008} are consistent with a nodeless order parameter. Single crystal measurements of the London penetration depth using a tunnel diode resonator (TDR) in SmFeAsO$_{1-x}$F$_x$  \cite{Malone08} and NdFeAsO$_{1-x}$F$_x$  \cite{Martin08} have both found exponential behavior in $\lambda(T \rightarrow 0)$. The superfluid density has been successfully fit to a single moderately anisotropic gap \cite{Martin08} and a two gap model without nodes \cite{Malone08}. However, the question of the order parameter symmetry in the 122 iron pnictides is still open. Over the past two decades much attention has been devoted to the cuprates, where the band structure is essentially two-dimensional and line nodes in the gap result in a $\lambda(T) \sim T$ behavior in the clean limit \cite{cuprates_review}, changing to $\sim T^2$ by the impurities \cite{Hirschfeld93}. Still, other materials may have nodes in a three-dimensional gap. There is evidence for point nodes in UBe$_{13}$ \cite{Einzel1986}, UPt$_3$ \cite{Brison2000}, PrOs$_4$Sb$_{12}$ \cite{Chia2003,MacLaughlin2008}, and possibly nonmagnetic borocarbides \cite{Izawa2002}.

In this Letter, we focus on the Ba(Fe$_{0.93}$Co$_{0.07}$)$_2$As$_2$  member of the 122 family, for which high quality single crystals are available \cite{Sefat2008a,Yamamoto2008,Prozorov2008}. The behavior in the mixed state is  similar to the high $ T_c$ cuprates, such as Y-Ba-Cu-O and Nd-Ce-Cu-O \cite{Prozorov2008}, and one could expect to find unconventional behavior in other properties as well.

\begin{figure}[tb]
\includegraphics[width=8.2cm]{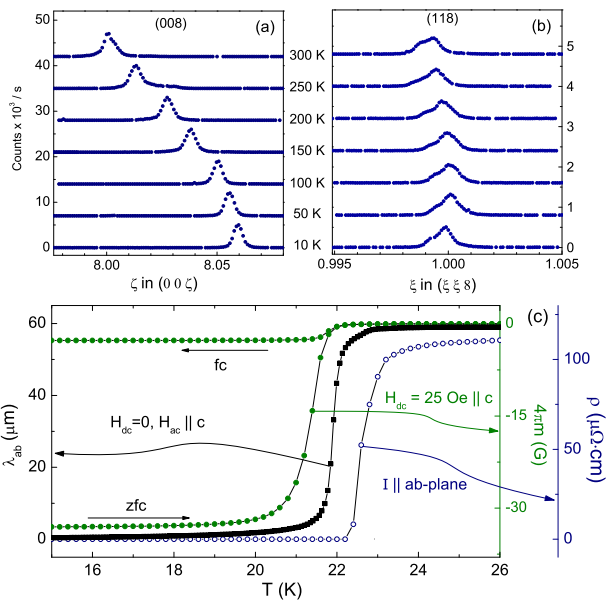}
\caption{(Color online)~Characterization of Ba(Fe$_{0.93}$Co$_{0.07}$)$_2$As$_2$  single crystals. (a) Longitudinal x-ray scans through the position of the (008) reflection for indicated temperatures. (b) similar ($\xi$ $\xi$ 0) scans through the position of the (118) reflection. (c) In-plane resistivity (open circles), dc magnetization (filled circles) and in-plane penetration depth  (squares).}
\label{fig1}
\end{figure}

The parent compound BaFe$_2$As$_2$ has been studied in detail elsewhere \cite{Nini08}. Single crystals of superconducting Ba(Fe$_{1-x}$Co$_x$)$_2$As$_2$ were grown out of FeAs flux using standard high temperature solution growth techniques. The FeAs and CoAs powders were mixed with Ba according to  the ratio Ba:FeAs:CoAs=1:3.6:0.4, placed into an alumina crucible, and then sealed in a quartz tube under an approximately 1/3 atm of argon gas. The sealed quartz tube was slowly heated to 1180 $^{\circ}$C, held for two hours, and then cooled to 1000 $^{\circ}$C over the course of 30 hours. Upon reaching 1000 $^{\circ}$C the FeAs was decanted from the single crystals. The crystal dimensions can be as large as $12 \times 8 \times 1$ mm$^3$. 

Crystals selected from several different batches have been extensively characterized by various techniques. Elemental analysis has shown that the actual Co concentration (averaged over six measurements) is Co/(Co+Fe)=7.4 \% $\pm 0.2$ \%. Powder x-ray diffraction on ground crystals has yielded tetragonal lattice constants of $a=3.9609 \pm 0.0008$ \AA\ and $c=12.9763 \pm 0.004$ \AA. Single crystal x-ray diffraction measurements have found no evidence of a tetragonal to orthorhombic structural transition, as shown in Fig.~\ref{fig1}(a)-(b), and the comparative analysis of the peak widths indicates homogeneous doping. Figure \ref{fig1} (c) shows in-plane resistivity, dc magnetization, and TDR penetration depth. $T_c$, as defined by zero resistivity, coincides with the onset of magnetization and is close to the shoe of the TDR data.

\begin{figure}[tb]
\includegraphics[width=8.2cm]{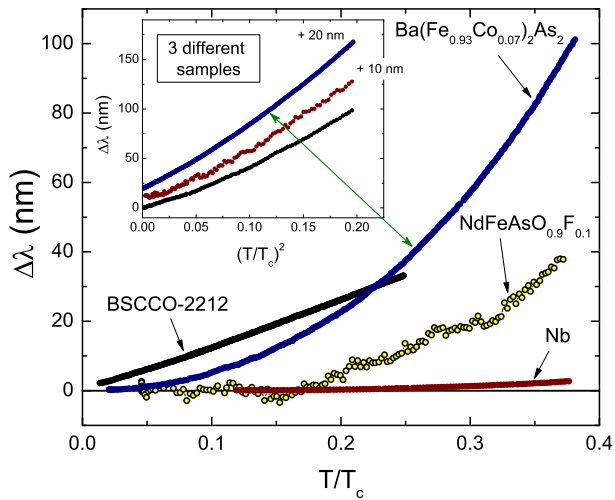}
\caption{(Color online) Comparison of the low temperature $\Delta \lambda$(T) measured in Ba(Fe$_{0.93}$Co$_{0.07}$)$_2$As$_2$  with the known d-wave (BSCCO) and s-wave (Nb) superconductors as well as NdFeAsO$_{1-x}$F$_x$. Inset: $\Delta\lambda$ vs. $(T/T_c)^2$ for three different samples of Ba(Fe$_{0.93}$Co$_{0.07}$)$_2$As$_2$  from different batches. (Curves are offset for clarity).}
\label{fig2}
\end{figure}

Measurements of the London penetration depth were performed using a tunnel diode resonator technique \cite{vandegrift,Prozorov00a,Prozorov00b}. A properly biased tunnel diode acts as an ac power source for the $LC$ circuit that promotes spontaneous oscillations at a natural frequency, $2 \pi f_0 = 1/\sqrt{L_0C}$, where $L_0$ is the inductance of the coil without a sample inside. The sample, mounted on a sapphire rod, is inserted into the coil without touching it. Throughout the measurement, the circuit is stabilized at 5.0 K $\pm$ 3 mK to ensure a stability in the resonance of less than 0.05 Hz over a period of several hours. In our setup $f_0 \approx 14$ MHz, which implies that the frequency resolution is better than 5 parts in 10$^9$. This translates into $\sim 1$ \AA\ resolution in $\Delta \lambda$ for mm-sized crystals.  For samples much smaller than the coil (small filling factor), the data are noisier,  as can be seen in the data for NdFeAsO$_{1-x}$F$_x$  in Fig.~\ref{fig2}. Due to the diamagnetic susceptibility $\chi$ of the sample, the inductance changes to a new value, $L$, and the resonant frequency shifts accordingly, $\Delta f=f-f_0 \approx - G4 \pi\chi (T)$, where $G=f_{0}V_{s}/2V_{c}\left(  1-N\right)$ is the calibration constant, $V_{s}$ is the sample volume, $V_{c}$ is the effective coil volume and $N$ is the demagnetization factor. For a superconductor in the Meissner state, $4 \pi \chi = \lambda/R \tanh{R/\lambda}-1$, where the effective dimension $R$ takes into account the actual sample shape \cite{Prozorov00a}. While in the coil, the sample experiences a $\sim$ 10 mOe ac excitation field, much smaller than the first critical field, $H_{c1}$, and therefore the London penetration depth is measured. The calibration constant $G$ has been determined by two different techniques \cite{Prozorov00a,Prozorov00b}. First, the sample was physically pulled from the coil at a low temperature to determine the full frequency change due to its presence. Second, the normal  state skin depth was calculated from the measured resistivity just above $T_c$, Fig.~\ref{fig1}(c). Both approaches yield the same result, which also rules out inhomogeneity of our samples. The TDR technique precisely measures changes in the penetration depth, but its absolute value is more difficult to obtain \cite{Prozorov00b}. To estimate $\lambda(0)$, we determined the lower critical field, $H_{c1}$, by measuring the $M(H)$ loops while increasing the maximum field until nonlinearity due to vortices appeared. Using $H_{c1}=\Phi_0/(4 \pi \lambda^2) (\ln{\lambda/\xi}+0.5)$, with a value for the coherence length of $\xi=2$ nm, as estimated from the upper critical field, we have obtained $\lambda (0) \approx 208$ nm. This value is similar to other reports, where $\lambda (0) \approx 254$ nm for La-1111 \cite{Luetkens2008} and 190 nm for Sm-1111 \cite{Drew2008} from $\mu$SR measurements. 

\begin{figure}[tb]
\includegraphics[width=8.2cm]{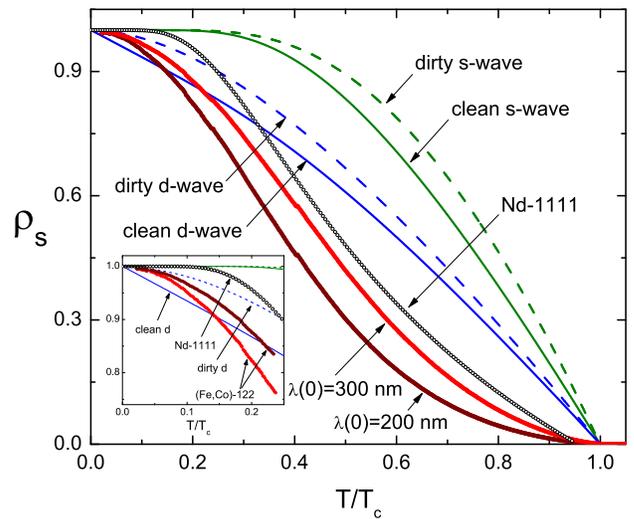}
\caption{(Color online) Superfluid density, $\rho_s$, as a function of normalized temperature for Ba(Fe$_{0.93}$Co$_{0.07}$)$_2$As$_2$  using $\lambda(0) = 200$ nm and $300$ nm.  Also shown for comparison are calculated curves for clean and dirty s-wave and d-wave pairings as well as best fit of NdFeAsO$_{1-x}$F$_x$ compound (circles) from Ref.\onlinecite{Martin08}.}
\label{fig3}
\end{figure}

The low temperature behavior of $\Delta\lambda (T) = \lambda(T)-\lambda(0)$ and the superfluid density, $\rho_s(T) \equiv \left(\lambda(0)/\lambda(T)\right)^2$, are commonly used to determine the symmetry of the superconducting pairing state \cite{Prozorov06}. In the case of a fully gapped Fermi surface, $\Delta\lambda(T)/\lambda(0) \approx \sqrt{\pi\Delta\left( 0\right)/2 k_B T}\exp{\left(-\Delta\left( 0\right)/k_BT\right)}$, which is also true for the cases of an anisotropic gap and two-gap superconductivity if one allows $\Delta(0)/T_c$ to be a free parameter. For the isotropic s-wave BCS case, this form is applicable for $T \leq T_c/3$. This is consistent with the data taken in our TDR system for the conventional BCS superconductor Nb, Fig.~\ref{fig2}. An $s-$wave behavior, albeit with an anisotropic gap with smaller amplitude, is found in NdFeAsO$_{1-x}$F$_x$~\cite{Martin08}, Fig.~\ref{fig2}. In the case of d-wave pairing, $\Delta\lambda(T)/\lambda(0) \approx T \left[ 2 \ln 2/\alpha \Delta(0)\right]$, where $\alpha  = \Delta ^{ - 1} \left( 0 \right)\left| {d\Delta \left( \phi  \right)/d\phi } \right|_{\phi  \to \phi _{node} }$  \cite{Xu95}. This linear behavior is observed in our measurements on clean BSCCO-2212 crystals, Fig.~\ref{fig2}. The above equations are applicable only for clean cases. Impurity scattering does not affect the isotropic s-wave gap but it does suppress the d-wave gap. The superfluid density, however, does change in both cases, as shown in Fig.\ref{fig3}. In the case of d-wave pairing, scattering leads to $\Delta \lambda(T) \sim  T^2$ \cite{Hirschfeld93}. For an s-wave superconductor in the dirty limit, $\rho_s(T)=\Delta(T)/\Delta(0)\tanh{[\Delta(T)/2 k_B  T]}$\cite{Tinkham96}.

The observed low-temperature behavior of $\lambda(T)$ in Ba(Fe$_{0.93}$Co$_{0.07}$)$_2$As$_2$  is clearly non-exponential, Fig.~\ref{fig2}. The inset of Fig.~\ref{fig2} shows the data plotted as a function of $(T/T_c)^2$ for samples from different batches all having different sizes and aspect ratios.  The middle curve corresponds to the smallest sample and therefore the smallest filling factor. The best fit to $\lambda(T) \sim T^n$ yields $n=2.4 \pm 0.1$, where the error reflects the fitting of different curves. For comparison, the penetration depth for the NdFeAsO$_{1-x}$F$_x$  crystal, plotted as open circles in Fig.~\ref{fig2}, is flat below about $0.2 T_C$, providing evidence for a fully gapped Fermi surface \cite{Martin08}.

To gain a better understanding of the pairing state, in Fig.~\ref{fig3} we plot $\rho_s (T)$ along with the known $s-$ and $d-$ wave behavior and the best fit for NdFeAsO$_{1-x}$F$_x$  \cite{Martin08}. The data for the Ba(Fe$_{0.93}$Co$_{0.07}$)$_2$As$_2$   samples are plotted  for $\lambda(0) = 200$ nm and $300$ nm. The inset of Fig.~\ref{fig3} zooms into the low temperature region showing that the observed non-exponential behavior persists down to $0.02 Tc$. As described above, the value of $\lambda(0) \approx 200$ nm was found from $H_{c1}$ and is consistent with the literature \cite{Drew2008,Luetkens2008}. The curve for $\lambda(0) = 300$ nm produces $\rho_s (T)$ closer to standard curves, but no $\lambda(0)$ can change the non-exponential low temperature behavior. At intermediate temperatures, the experimental $\rho_s(T)$ shows significant departure from the d-wave and s-wave curves, Fig.~\ref{fig3}. Possible reasons for this behavior could be significant gap anisotropy or multi-gap pairing with different gap amplitudes. We note that for NdFeAsO$_{1-x}$F$_x$  crystals, a similar behavior in the intermediate temperature region has been found (Fig.~\ref{fig3} open circles) and explained in terms of an anisotropic s-wave gap \cite{Martin08}. However, this does not lead to the low-temperature $T^2$ behavior.

\begin{figure}[tb]
\includegraphics[width=6.2cm]{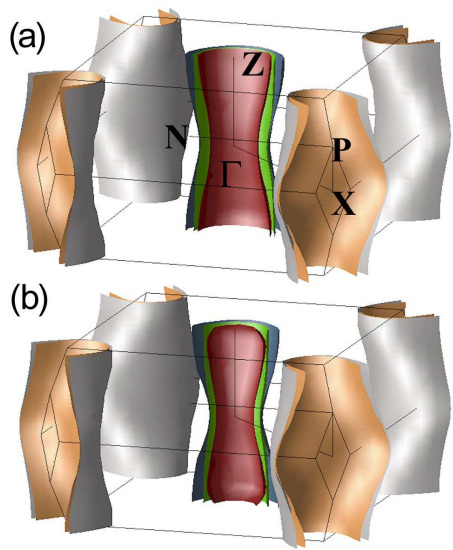}
\caption{(Color online) (a) Fermi surface of non-superconducting BaFe$_2$As$_2$   and (b) Ba(Fe$_{0.93}$Co$_{0.07}$)$_2$As$_2$ with experimental As positions. Note significant enhancement of a 3D character around $\Gamma$, $Z$ and $X$.}
\label{fig4}
\end{figure}

A possible explanation for the unusual exponent in $\lambda \sim T^n$ with $n \geq 2$, is the existence of a three-dimensional (3D) gap with point nodes. In such a case, a quadratic behavior of $\lambda(T)$ is predicted and observed in the clean case \cite{Einzel1986,Chia2003}. With impurities, the exponent $n$ increases, consistent with the observed $n=2.4$. Similarly, in the 2D d-wave case $T-$linear behavior of the penetration depth changes to $T^2$ \cite{Hirschfeld93}. Since a three-dimensional portion of the Fermi surface is a crucial ingredient for the point-node scenario, electronic structure calculations, Fig.~\ref{fig4}, were performed within the full-potential linearized plane wave (FLAPW) approach \cite{LAPW} using the local density approximation (LDA) \cite{LDA}. Both relaxed and experimental As positions were used with the experimental lattice constants for the BaFe$_2$As$_2$ \cite{exp_BaFe2As2} and Ba(Fe$_{0.93}$Co$_{0.07}$)$_2$As$_2$ (see above). A virtual crystal approximation was employed with Fe replaced by a virtual atom having Z=26.07 to simulate 7.4\% Co doping. The Fermi velocities, calculated using the BolzTrap \cite{BolzTrap} package, are very sensitive to the As positions, a situation similar to calculations of magnetic properties \cite{Mazin_Singh,Singh_BaFe2As2}. The calculated coordinate, z$_{As}$=0.341, is comparable to previous calculations, z$_{As}$=0.342 \cite{Singh_BaFe2As2}, but is significantly lower than the experimental value of z$_{As}$=0.355. This downshift by 0.16 \AA\ increases the band dispersion along the z direction. The calculations with relaxed z$_{As}$ give $\gamma^2_\lambda = \left< v^2_x \right>/\left< v^2_z \right>=3$ for pure compound and $\gamma^2_\lambda = 2.7$ for Ba(Fe$_{0.93}$Co$_{0.07}$)$_2$As$_2$. With the experimental z$_{As}$, the difference is much larger, $\gamma^2_\lambda = 12.1$ and 9.0, for pure and doped materials, respectively. 

The fact that the superconducting gap functions in NdFeAs(O$_{1-x}$F$_{x}$)
and Ba(Fe$_{0.93}$Co$_{0.07}$)$_2$As$_2$  behave so differently may imply that the two systems are governed by truly distinct pairing mechanisms. However, both systems share the Fe-As planes as key structural building blocks. Electronic structure calculations demonstrate that the states at the Fermi energy are dominated by Fe-$3d$ states. Thus, we discuss the implications of our findings under the premise that there is a common pairing mechanism and a common pairing symmetry in both materials. Our earlier experimental results for NdFeAsO$_{1-x}$F$_x$  rule out a $d$-wave pairing state. Previously, we argued that a likely interpretation of the moderate anisotropy of the superconducting gap, $\Delta \left( \varphi \right)$, \cite{Martin08,Liu08} can naturally be understood as a result of an electronic interband pairing interaction \cite{Mazin2008} that changes the sign of the gap. In this scenario, anisotropic $s$-wave pairing occurs with the nodes located between Fermi surface sheets \cite{Rastko2008}. Then, $\Delta \left( \varphi \right) \neq 0$ for all $\varphi$ and is slightly smaller along the diagonals of the Brillouin zone connecting $\Gamma $ and $X$ points. As shown in \ref{fig4}, the Fermi surface sheets of Ba(Fe$_{0.93}$Co$_{0.07}$)$_2$As$_2$  are more 3D than those of the parent BaFe$_{2}$As$_{2}$, and even more so than NdFeAs(O$_{1-x}$F$_{x}$) \cite{Martin08}. The ``bellies'' of the sheets around the $X-$points in the Brillouin zone corners reach out toward the nodes and a vanishing gap can easily occur at a point or small ellipse close to the ``belly button''. As in NdFeAs(O$_{1-x}$F$_{x}$), this reduction of the gap is expected to occur along the directions connecting the  $\Gamma$ and $X$ points. Such an anisotropic $s$-wave state \textit{with} nodes implies that the existence of nodes depends sensitively on the Fermi surface shape, in sharp contrast to $d$- or $p$-wave states where the nodes are enforced by the symmetry.  
 
We thank V.~G.~Kogan, J.~R.~Clem, and A.~Kaminski for discussions and comments. Work at the Ames Laboratory was supported by the Department of Energy-Basic Energy Sciences under Contract No. DE-AC02-07CH11358. R. P. acknowledges support from Alfred P. Sloan Foundation.

 \end{document}